\documentclass[nojss]{jss}
\usepackage{amsmath, amssymb}
\usepackage{booktabs, array}

\usepackage{rotating}
\usepackage{placeins}

\author{Cheolju Kim\\Boston University \And Zhongjun Qu\\Boston University \And Pierre Perron\\Boston University}
\title{\pkg{sysbreak}: An \proglang{R} Package for Structural Breaks in a System of Equations}
\Plainauthor{Cheolju Kim, Zhongjun Qu, Pierre Perron}
\Plaintitle{sysbreak: An R Package for Structural Breaks in a System of Equations}
\Shorttitle{\pkg{sysbreak}: Structural Breaks in a System of Equations}

\Abstract{The \pkg{sysbreak} package provides methods for detecting, estimating,
and conducting inference on multiple structural breaks in a linear system
of equations, such as multivariate linear regressions, seemingly
unrelated regressions, vector autoregressions, and network vector
autoregressions. Breaks can occur in the regression coefficients,
the covariance matrix of the errors, or both, and can be restricted to a subset of
the coefficients. The main function, \code{sysbreak()}, returns the number of breaks detected and
confidence intervals for the break dates and the regression
coefficients.
A separate function \code{ordbreak()}
estimates ordered breaks, in which a break in one subset of equations
occurs before a break in another. Specialized functions for vector
autoregressions (\code{varbreak()}) and network vector autoregressions (\code{netbreak()}) are also provided. This paper outlines
the underlying methods and illustrates the use of these functions through
three examples: U.S. house prices and the macroeconomy,  international inflation dynamics and spillover,  and a network vector
autoregression with simulated data. All datasets are
included in the package,  available on CRAN.}

\Keywords{structural break, systems of equations, multivariate regression, vector autoregression, network vector autoregression}
\Plainkeywords{structural break, systems of equations, multivariate regression, VAR, network VAR}

\Address{
  Cheolju Kim, Zhongjun Qu, Pierre Perron\\
  Department of Economics\\
  Boston University\\
  270 Bay State Road\\
  Boston, MA 02215, United States of America\\
  E-mail: \email{qu@bu.edu} (Zhongjun Qu)
}

\begin{document}

\section{Introduction}

The issue of structural change in economic, financial and many other types of time series has
been extensively studied. Much of the literature has focused on
single-equation models, but structural change frequently affects systems of
equations rather than a single relationship. For example, in macroeconomic vector
autoregressions, a shift in the monetary policy rule may be
accompanied by shifts in the response equations for output and
inflation. In cross-country inflation studies, the conditional-mean
coefficients and the disturbance covariance may both exhibit structural breaks (Bataa, Osborn, Sensier, and van Dijk, 2013). In financial applications, sovereign bond
spread correlations may rise abruptly during contagion episodes (Blatt,
Candelon, and Manner, 2015). Treating equations one at a time in such
settings misses cross-equation mechanisms. It may also lead to less efficient estimates of break dates.

Qu and Perron (2007) developed a general framework for structural
change in a linear system of equations. It allows common breaks
across all equations, breaks in a subset of equations, and partial breaks
in a subset of coefficients. The tests can detect breaks in the
coefficients, in the covariance matrix of the errors, or in both. The framework also covers locally ordered
breaks, where a break in one set of equations (e.g., a monetary
policy rule) is followed by a break in another (e.g., the behavioral
equations). The  ordering helps identify policy lags, which makes the model useful in a causal effects setting.

Key aspects of these methods are the following. The break dates
are estimated jointly with the regime-specific coefficient vectors and
covariance matrices, as the global maximizer of a Gaussian quasi-likelihood
over admissible partitions of the sample. Dynamic programming is used to
reduce this computation to at most $O(T^2)$ operations for any number of
breaks, where $T$ is the sample size. The resulting break-fraction estimates
converge at a faster rate than the coefficient estimates. Their limiting
distribution leads to confidence intervals that do not require simulations.
The presence of a break and the number of breaks are then determined using
likelihood-ratio statistics. Robust alternatives based on supremum-type Wald
statistics are available when the disturbances are heteroskedastic or
serially correlated. Finally, the critical values for these tests are
computed from response surfaces following Bai and Perron (2003b),
as functions of the number of parameters allowed to change, the number of
breaks, and the trimming parameter.

To date, there has been no published \proglang{R} package that implements the Qu and
Perron (2007) framework in an easy-to-use fashion. This led us to develop the
\pkg{sysbreak} package, which is available on CRAN. The package provides
four estimation functions organized around the specifications supported:
\begin{itemize}
\item \code{sysbreak()}: multivariate regressions and seemingly unrelated
  regression systems, specified by a formula or a list of formulas.
\item \code{varbreak()}: vector autoregressions, with the intercept and lagged
  values built automatically from the data.
\item \code{netbreak()}: network vector autoregressions with a fixed
  network matrix $W$, allowing breaks in the network channel, the
  own-dynamics channel, or both.
\item \code{ordbreak()}: models with two ordered breaks, where a break in
  one subset of equations is followed by a break in another.
\end{itemize}
The package also provides two functions that further analyze a fitted model
(Section~\ref{sec:extensions}). The function \code{brtype()} tests whether
the volatilities, the correlations, and the coefficients change at each
break. This helps assess the nature of each change. The function
\code{dyspill()} computes the spillover measures of Diebold and Yilmaz
(2012) in each regime, useful for studying issues such as contagion.
These two functions and the network vector autoregression extend the
original framework of Qu and Perron (2007). 

We illustrate the package through three applications: 
\begin{itemize}
\item The first uses \code{sysbreak()} to revisit the study of house prices
and the U.S. macroeconomy by Ahamada and Diaz Sanchez (2013). The model is a
three-variable system for output, house prices, and the federal funds rate.
A break is found in 1982 when all coefficients are allowed to change, and in
1985 when only the house price coefficients are allowed to change. Granger
causality tests show that house prices Granger-cause output only in the
second regime.

\item The second uses \code{varbreak()} to revisit the international
inflation dynamics as in Bataa et al.~(2013), with
a four-country vector autoregression. The methods detect coefficient breaks
in 1983 and 1991, and an additional covariance break in 1999. The function
\code{brtype()} shows that the first two breaks represent changes in the
volatilities and the coefficients, while the third affects the volatilities
and the correlations. Furthermore, \code{dyspill()} shows that the spillover
index between the countries doubles at the third break, while it is not affected by the first
two breaks. This suggests that inflation shocks have become more
synchronized across countries since 1999, consistent with closer integration
of these economies.

\item The third applies \code{netbreak()} to a network vector autoregression
with two breaks. The data contain 20 units and 300 observations, simulated
from the data generating process of Han and Lee (2024). Both breaks are estimated at the true
dates when all coefficients are allowed to change. As a robustness check, we
then hold the intercepts constant across regimes, a restriction that is not
satisfied in the data. We find that the estimated break dates remain the
same, while the coefficient estimates exhibit distortions. This illustrates
the different effects of a false restriction on break date and coefficient
estimation.

\end{itemize}

Several \proglang{R} packages are available which deal with structural breaks in a single
equation. \pkg{strucchange} (Zeileis, Leisch, Hornik, and Kleiber, 2002;
Zeileis, Kleiber, Kr\"amer, and Hornik, 2003) provides fluctuation and
$F$ tests and dates multiple breaks in a linear regression using the algorithm
of Bai and Perron (2003a). \pkg{mbreaks} (Nguyen, Yamamoto, and Perron,
2024) implements the procedures of Bai and Perron (1998) for a single
equation, including the sequential test and the confidence intervals for
the break dates. However, neither handles a system of equations. Packages for
change points in multivariate series, such as \pkg{ecp} (James and
Matteson, 2015), detect changes in the distribution of a vector series
without a regression structure. The methods of Qu and Perron (2007) have been implemented in
\proglang{GAUSS}, with a direct \proglang{MATLAB} translation. Relative to these implementations, the new
\proglang{R} package provides a substantially simplified user interface,
where the model and the coefficients allowed to change are specified by
simple formulas or keywords.
In addition, it introduces the functions \code{varbreak()} and \code{netbreak()} for vector
autoregressions and network vector autoregressions. It also provides
tools for analyzing the fitted model, including \code{brtype()} for
identifying what changes at each break and \code{dyspill()} for
analyzing spillovers across regimes. These features are illustrated in the empirical applications.

The framework is designed for settings where the sample size $T$ is large relative to the number of equations $n$.
It is not intended for panel data with a large cross-section. Structural
change in panels with many units has been studied in a branch of
the literature that typically relies on shrinkage-based estimation; see, for example,
Qian and Su (2016), Li, Qian, and Su (2016), and Okui and Wang
(2021). A separate literature addresses structural change in factor
models for large panels; see, for example, Breitung and
Eickmeier (2011), Chen, Dolado, and Gonzalo (2014), and Han and Inoue
(2015). These settings are not covered by the present package.

The rest of the paper is organized as follows. Section~\ref{sec:methods}
outlines the methods. Section~\ref{sec:package}
describes the package: its functions, the arguments of \code{sysbreak()},
and the output. Section~\ref{sec:applications} uses the
package to investigate the three examples mentioned above. Section~\ref{sec:conclusion}
concludes.

\section{Methods for estimation and inference}
\label{sec:methods}

This section outlines the following aspects: (1) the multivariate structural change model and
the econometric issues of interest; (2) the main steps for estimating the
break locations and regression parameters when all coefficients are
allowed to change across regimes; (3) the main steps when a partial break
restriction is imposed, meaning that a subset of the coefficients is held
constant across regimes; (4) tests for the presence and number of breaks; and
(5) inference on the break dates and on the regression coefficients.

\subsection{The model and issues of interest}

Let $T$ be the sample size and assume that the system has $m$ structural
breaks at unknown dates $T_1^0 < T_2^0 < \cdots < T_m^0$, giving $m+1$
regimes. Let $y_{it}$ and $x_{it}$ be the dependent variable and the
$p_i \times 1$ vector of regressors associated with the $i$th equation,
$i = 1, \ldots, n$. Then the system in the $j$th regime, defined by
$T_{j-1}^0 < t \le T_j^0$ for $j = 1, \ldots, m+1$ (with the convention
that $T_0^0 = 0$ and $T_{m+1}^0 = T$), is given by
\begin{equation}
\begin{aligned}
y_{1t} &= x_{1t}' \beta_{1,j}^0 + u_{1t},\\
&\;\;\vdots\\
y_{nt} &= x_{nt}' \beta_{n,j}^0 + u_{nt},
\end{aligned}
\label{eq:model-eqwise}
\end{equation}
where $u_t = (u_{1t}, \ldots, u_{nt})' \sim (0, \Sigma_j^0)$ and
$\Sigma_j^0$ is the $n \times n$ disturbance covariance in regime $j$. It
can be held constant across regimes or allowed to change at every
break. 
Note that different equations can contain common as well as different regressors,
and the numbers of regressors for different equations can also differ.
In matrix notation, the system at time $t$ is
\[
y_t = x_t' \beta_j^0 + u_t,
\]
with $y_t = (y_{1t}, \ldots, y_{nt})'$,
$x_t' = \mathrm{diag}(x_{1t}', \ldots, x_{nt}')$, and
$\beta_j^0 = (\beta_{1,j}^{0\prime}, \ldots, \beta_{n,j}^{0\prime})'$ of
dimension $p = \sum_i p_i$. Writing this out explicitly for the $m+1$ regimes, we have
\begin{equation}
y_t =
\begin{cases}
x_{t}' \beta_1^0 + u_t, & t = 1, \ldots, T_1^0,\\
x_{t}' \beta_2^0 + u_t, & t = T_1^0 + 1, \ldots, T_2^0,\\
\vdots & \vdots \\
x_{t}' \beta_{m+1}^0 + u_t, & t = T_m^0 + 1, \ldots, T,
\end{cases}
\qquad u_t \sim (0, \Sigma_j^0).
\label{eq:model}
\end{equation}
When using the package, users do not need to construct $x_t$. Instead, the model is specified by a formula or a list of formulas, one for each equation, and $x_t$ is constructed internally.

The implementation allows partial structural change. Users can select a subset of the regression coefficients that are allowed to change across regimes, and the remaining coefficients are then constrained to be equal across regimes. The selection can differ across equations. For example, all coefficients in the first equation are allowed to change, the first two coefficients in the second equation are allowed to change, and all coefficients in the third equation remain invariant across regimes. Qu and Perron (2007) allow more general restrictions of the form $g(\beta, \mathrm{vec}(\Sigma)) = 0$, where $\beta$ and $\Sigma$ contain the coefficients and the covariance matrices of all regimes. The implementation focuses on partial structural change because it covers many empirical settings and can be specified through a simple interface. The following examples illustrate the class of specifications supported.

\medskip
\textit{Example 1.} Bai, Lumsdaine, and Stock (1998) considered common breaks across a system of macroeconomic variables and showed that information from ancillary equations improves the precision of the break-date estimator. Such a system can be estimated with \code{sysbreak()}. To see how the interface works in this case, consider a simple two-equation example with two common regressors (more information on the implementation is given in Section~\ref{sec:package}). The model is then specified by the formula \code{cbind(y1, y2) \textasciitilde{} x1 + x2}. If instead the regressors differ across equations, as in seemingly unrelated regressions, a list of formulas can be supplied, one for each equation. For example, \code{list(y1 \textasciitilde{} x1 + x2, y2 \textasciitilde{} x3)} specifies a system in which the first equation has regressors \code{x1} and \code{x2} and the second has only \code{x3}. In both cases an intercept is included in each equation unless the formula excludes it, as in \code{y2 \textasciitilde{} 0 + x3}.

\medskip
\textit{Example 2.} Bai (2000) studied structural change in vector autoregressive models in which the break may occur in the coefficients, the disturbance covariance, or both. Such a model can be estimated with \code{varbreak()}. To illustrate, consider a VAR(1) for two variables, stored as the columns \code{y1} and \code{y2} of \code{y}. Then, one can use \code{varbreak(y, lags = 1, m = 3, brv = FALSE)} to allow up to three breaks, with all coefficients changing and the covariance held constant. Setting \code{brv = TRUE} will allow the covariance to change as well. Additional features related to partial structural change and the computation of spillovers are discussed in Section~\ref{sec:applications}. 

\medskip
\noindent
\textit{Example 3.} Zhu, Pan, Li, Liu, and Wang (2017) introduced the network vector autoregression (NVAR), in which the dynamics of each unit depend on its own past and on a weighted average of neighboring units through a known network matrix $W$. For an NVAR of order one with structural breaks, the model is
\[
y_t = \mu_j + (\alpha_j W + \gamma_j I) y_{t-1} + u_t
\quad \text{for } T_{j-1}^0 + 1 \le t \le T_j^0,
\]
so that each regime is characterized by three groups of parameters: $\mu_j$ (the intercepts), $\alpha_j$ (dependence through the network channel), and $\gamma_j$ (dependence through the own-dynamics channel). Such a model can be estimated with \code{netbreak()}. Users provide the network matrix $W$ and the lag order, and choose which of the three groups of parameters are allowed to change across regimes. More details are discussed in Section~\ref{sec:app-network}.

\medskip
\noindent
\textit{Example 4.} Qu and Perron (2007, \S 5) introduced locally ordered breaks, in which a break in one subset of equations (e.g., a monetary policy rule) is followed by a break in another subset (e.g., a Phillips curve), where the interval between the two is short relative to the sample size. Such a model can be estimated with \code{ordbreak()}. The function estimates the two break dates, $k_1$ and $k_2$, jointly subject to the restriction $k_1 \le k_2$. It constructs confidence intervals by simulating the limiting distribution in Qu and Perron (2007, Theorem 4). It also computes the sup-likelihood-ratio test of Qu and Perron (2007, \S 6.4) for no break against locally ordered breaks, over pairs of dates whose distance does not exceed a user-specified threshold. The critical values are those of a single-break test, with the number of parameters equal to the total number of coefficients allowed to change at either of the two dates. The test for globally ordered breaks (Qu and Perron, 2007, \S 6.5) is left for a future version.

\subsection*{Issues of interest}

The methods in the package address the following issues:
\begin{enumerate}
\item Estimation of the break locations and regression parameters when
  the number of breaks is known and all coefficients are allowed to change
  across regimes.
\item Estimation of the break locations and regression parameters when
  the number of breaks is known and a partial break restriction is
  imposed. 
\item Determination of the number of breaks through likelihood-ratio and HAC-robust tests, together with the
  WDmax test (a test for an unknown number of breaks up to some pre-specified maximum) and a sequential $l+1$-against-$l$ break test.
\item Confidence intervals for the estimated break dates from the
  shrinking-shift limiting distribution.
\item Standard errors and confidence intervals for the regime-specific
  regression coefficients, conditional on the estimated break dates.
\end{enumerate}
The main functions of the package perform all of these tasks in a single
call. In what follows, we describe each step in turn.

\subsection{Estimating break locations and parameters when all coefficients change}
\label{sec:est-unrestricted}
Suppose the number of breaks $m$ is known and all coefficients are allowed
to change across regimes. Given a candidate partition
$\mathcal{T} = (T_1, \ldots, T_m)$, the Gaussian quasi log-likelihood is
\begin{equation}
\ell_T(\mathcal{T}, \beta, \Sigma) =
-\frac{Tn}{2} \log(2\pi)
-\frac{1}{2} \sum_{j=1}^{m+1} \bigl[
(T_j - T_{j-1}) \log \lvert \Sigma_j \rvert
+ \sum_{t=T_{j-1}+1}^{T_j} (y_t - x_t' \beta_j)' \Sigma_j^{-1}
(y_t - x_t' \beta_j) \bigr],
\label{eq:loglik}
\end{equation}
with the convention $T_0 = 0$ and $T_{m+1} = T$. The quasi-maximum
likelihood estimator solves
\begin{equation}
(\hat{\mathcal{T}}, \hat{\beta}, \hat{\Sigma})
= \arg\max_{\mathcal{T} \in \Lambda_\varepsilon,\ \beta,\ \Sigma}
\ell_T(\mathcal{T}, \beta, \Sigma),
\label{eq:qmle}
\end{equation}
where $\Lambda_\varepsilon$ is the set of admissible partitions in which
each regime has length at least $\varepsilon T$:
\[
\Lambda_\varepsilon = \{ (T_1, \ldots, T_m):
T_j - T_{j-1} \ge \varepsilon T\ (j = 1, \ldots, m+1),\
T_1 \ge \varepsilon T,\ T_m \le (1-\varepsilon) T \}.
\]
The trimming $\varepsilon$ is set through
\code{trim}; values between $0.10$ and $0.20$ are recommended.

The computation is carried out in two nested loops. In the inner loop,
for every candidate segment $(i, j)$ with $i \le j$, the Gaussian
quasi-likelihood is maximized over $(\beta_{(i,j)}, \Sigma_{(i,j)})$ using
the iterative feasible generalized least squares (FGLS) procedure of
Tobing and McGilchrist (1992). Because the segment $(i, j+1)$ differs
from $(i, j)$ by one observation, the FGLS estimates can be
updated recursively without recomputing the full segment likelihood. In the outer loop, a
dynamic-programming algorithm as in Bai and Perron (2003a) searches over
all admissible partitions to find the combination of $m+1$ segments
that maximizes the total likelihood. The overall computation is at most $O(T^2)$ for any number of breaks allowed. 

Under the assumptions of Qu and Perron (2007), the break-fraction
estimates $\hat{\lambda}_j = \hat{T}_j / T$ are consistent for
$\lambda_j^0 = T_j^0 / T$, and the regime-specific coefficient
and covariance estimates are $\sqrt{T}$-consistent with limiting
distributions that coincide with the known-break case. 

\subsection{Estimating break dates and parameters under a partial break restriction}
\label{sec:est-restricted}

When a partial break restriction is imposed, the break-date search cannot
be reduced to a triangular matrix of likelihood segments because the
restriction implies parameters that interact across regimes. Qu and Perron (2007,
\S 4.2) proposed the following iterative procedure, 
implemented in the package:

\begin{enumerate}
\item[1.] Estimate the break dates, assuming that all coefficients are allowed to change, by the dynamic-programming
  procedure of Section \ref{sec:est-unrestricted}.
\item[2.] Given the break-date estimates, estimate the
  restricted regime-specific coefficients and covariances by feasible
  GLS under the reparameterization  $\beta = R\theta$, where $\beta = (\beta_1', \ldots, \beta_{m+1}')'$ contains the coefficients in (\ref{eq:model}) for all regimes, $R$ is a known matrix of constants, and $\theta$ is a vector of unrestricted parameters. The package constructs $R$ internally.
\item[3.] Given the restricted coefficients and covariance estimates,
  update the break dates via a dynamic-programming search on the
  restricted likelihood segments.
\item[4.] Repeat Steps 2 and 3 until the break dates and the
parameter estimates stabilize.
\end{enumerate}
The user specifies which coefficients are allowed to change through the
argument \code{coef\_breaks}. Setting \code{coef\_breaks = "all"} lets all
coefficients change, as discussed in
Section~\ref{sec:est-unrestricted}. Setting \code{coef\_breaks = "none"}
allows no coefficient to change and is used together with
\code{brv = TRUE} to allow covariance breaks only. Supplying a list of
formulas such as
\code{list(y1 \textasciitilde{} x1, y2 \textasciitilde{} 0)} restricts
the breaks to apply to the coefficients named on the right-hand side of
each formula.

The iterative procedure is not guaranteed to reach the global optimum
because the dynamic programming step in the first iteration uses the unrestricted
likelihood as a starting point, though we found convergence to a local minimum to be rare in practice. To reduce the sensitivity of the
restricted estimator to the starting partition, the package provides an
optional bootstrap-restart step through the arguments
\code{restart = "brbp"} (resampling bootstrap restart) and
\code{restart = "brbp\_residual"} (residual bootstrap restart), following
the bootstrap restarting of Wood (2001). At each restart, the break dates are estimated on a bootstrap sample and are then used as the starting partition for the iterative procedure on the original data. The estimates are updated whenever the log-likelihood on the original data increases. The number of restarts is set by \code{restart\_B}.

\subsection{Testing for structural change}
\label{sec:testing}

The package computes four families of tests, following Qu and Perron
(2007, \S 6).

\subsubsection*{The SupLR test}

For a specified maximum number of breaks $m$, the SupLR test compares
the maximized log-likelihood under $m$ breaks with that under no break:
\begin{equation}
\text{SupLR}_T(m) = 2 \bigl[
\max_{\mathcal{T} \in \Lambda_\varepsilon} \ell_T(\mathcal{T}, \hat{\beta}, \hat{\Sigma})
- \ell_T(\emptyset, \tilde{\beta}, \tilde{\Sigma}) \bigr].
\label{eq:suplr}
\end{equation}
Here, for a given partition $\mathcal{T}$, $\hat{\beta}$ and $\hat{\Sigma}$ are the quasi-maximum likelihood estimates of the coefficients and the covariance under $m$ breaks, and $\tilde{\beta}$ and $\tilde{\Sigma}$ are the corresponding estimates under no break, computed from the full sample. Under the null hypothesis of no break and the assumptions of Qu and Perron
(2007), the limiting distribution of \eqref{eq:suplr} is a functional of
independent Wiener processes that depends only on the number of parameters
allowed to change, the number of breaks, and the trimming $\varepsilon$
(their Corollary 2). The package computes the critical values from response surfaces, following the approach of Bai and Perron (2003b). When the covariance is allowed to change, these critical values are computed assuming Gaussian errors. Qu and Perron (2007) allow for more general errors, but the limiting distribution is then not free of nuisance parameters. When only the coefficients are allowed to change, the critical values are valid without the Gaussian assumption. The same holds for the double-maximum and sequential tests described below.

\subsubsection*{Double-maximum tests}

When the true number of breaks is unknown, Qu and Perron (2007) recommend
the WDmax tests of Bai and Perron (1998):
\[
\text{WDmax}_T(M) = \max_{1 \le m \le M}
\frac{c_1(1, \alpha)}{c_1(m, \alpha)} \text{SupLR}_T(m),
\]
where $c_1(m, \alpha)$ is the asymptotic critical value of the $\text{SupLR}_T(m)$ test at significance level $\alpha$.
These tests detect the presence of some number of breaks in
$\{1, \ldots, M\}$ without requiring the user to specify the true
number of breaks.

\subsubsection*{Sequential test}

Given a partition with $l$ estimated breaks, the sequential test of
$l+1$ against $l$ breaks, denoted $\text{SEQ}_T(l+1|l)$, inserts one
additional break into each of the $l+1$ segments and takes the maximum of
the corresponding one-break SupLR statistics. The number of breaks is
determined by increasing $l$ until $\text{SEQ}_T(l+1|l)$ fails to
reject. In this package, each statistic is scaled by the Bartlett
(1938) factor $(T_s - b)/T_s$, where $T_s$ is the length of the segment,
$b = k + (n - p + 1)/2$, $k$ is the number of coefficients per equation in
the one-break model, and $p$ is the number of coefficients per equation allowed to
change (Anderson, 2003, Section 8.5). This small-sample correction matters
when $n$ is not small relative to $T_s$.

\subsubsection*{Robust versions}

When the disturbances $u_t$ are conditionally heteroskedastic or serially
correlated, the SupLR test is not asymptotically pivotal. Qu and Perron
(2007, Eq.~20) develop a robust SupF statistic based on a consistent estimate
of the long-run covariance of the scores (Andrews, 1991). The limiting distribution and critical values remain the same. The robust tests are available for coefficient breaks but
not for covariance-break tests since the critical values there are case-specific. It returns a missing value
with a warning when the latter is requested.

\subsubsection*{Recommended procedure}

We recommend the following procedure to determine the number of breaks:
\begin{itemize}
\item[Step 1.] Apply the WDmax test with a maximum number of breaks $M$
  (e.g., $M =3$). If it fails to reject, conclude that there is
  no break.
\item[Step 2.] If WDmax rejects, apply $\text{SEQ}_T(l+1|l)$
  starting at $l = 1$. Increase $l$ each time the test rejects,
  reestimating the break dates as the maximizer of the likelihood at
  each stage.
\item[Step 3.] Let $\hat{l}$ denote the first value for which
  $\text{SEQ}_T(l+1|l)$ fails to reject. Estimate the model with $\hat{l}$
  breaks and report break dates, coefficients, and confidence intervals.
\end{itemize}
This procedure controls the overall significance level from above at the
level used in the sequential test. This procedure is implemented in the package.

\subsection{Confidence intervals for the break dates and inference on the coefficients}
\label{sec:ci}

For each estimated break date $\hat{T}_j$, the shrinking-shift asymptotic
framework of Qu and Perron (2007, Theorem 3) gives
\begin{equation}
\hat{L}_j (\hat{T}_j - T_j^0)
\overset{d}{\longrightarrow} \arg\max_u Z_j(u),
\label{eq:break-ci}
\end{equation}
where $Z_j(u)$ is
a function of a two-sided Brownian motion with drift, and $\hat{L}_j$ is computed from the estimated changes in the coefficients and in the covariance matrix at the $j$th break, using the expressions in Qu and Perron (2007, pp.~473--474). As shown in Bai (1997), the limiting distribution has an analytical
cumulative distribution function; simulation is not needed.

Given the estimated break dates, the regime-specific coefficient
estimates and their standard errors are computed conditional on
$(\hat{T}_1, \ldots, \hat{T}_m)$. Because the break-fraction estimates
converge at a fast rate while the coefficient estimates converge at rate
$\sqrt{T}$, the sampling variability of the estimates of the break dates does not affect
the asymptotic distribution of the coefficient estimates, and therefore standard
$\sqrt{T}$-normal confidence intervals apply. The package returns the
regime-specific coefficient estimates and their standard errors through the \code{estimation} part of the
returned object.

\subsection{Extensions: Further analysis across and within regimes}
\label{sec:extensions}

After the break dates are estimated, it is often of interest to examine what has changed at each break, and how shocks are transmitted across equations within each regime. The package provides two functions, \code{brtype()} and \code{dyspill()}, for these purposes. Both functions take a fitted model as input and treat the estimated break dates as known. This is justified asymptotically because the break fractions are estimated at a fast rate.

\subsubsection*{Examining the type of change at each break}

For a break at $\hat{T}_j$, the function \code{brtype()} tests whether the residual volatilities, their correlations, and the regression coefficients differ between regimes $j$ and $j+1$, with $T_a$ and $T_b$ observations in the two regimes. Write $\Sigma_j^0 = S_j R_j S_j$, where $S_j$ is the diagonal matrix of standard deviations of the residuals and $R_j$ their correlation matrix. The three tests are described below.

\begin{itemize}
\item \emph{Volatility.} The test uses a two-sample Hotelling statistic on the squared
  residuals for the null hypothesis $S_j = S_{j+1}$:
  \[
  \frac{T_a T_b}{T_a + T_b} (\bar{s}_a - \bar{s}_b)' V^{-1}
  (\bar{s}_a - \bar{s}_b),
  \]
  where $\bar{s}_a$ and $\bar{s}_b$ are the regime means of
  $(\hat{u}_{1t}^2, \ldots, \hat{u}_{nt}^2)'$, and $V$ is their pooled
  covariance. It is compared with the $\chi^2_n$ distribution.
\item \emph{Correlation.} The test is that of Jennrich (1970) for the null hypothesis $R_j = R_{j+1}$, compared with the $\chi^2_{n(n-1)/2}$ distribution. It assumes
  normal errors.
\item \emph{Coefficients.} The test uses a Wald statistic for the null hypothesis $\beta_j^0=\beta_{j+1}^0$, compared with the $\chi^2$
  distribution.
\end{itemize}

\subsubsection*{Measuring spillovers within each regime}

For fitted models from \code{varbreak()} and \code{netbreak()}, the function
\code{dyspill()} computes the spillover measures of Diebold and Yilmaz
(2012) in each regime.  Let $\Psi_{s,j}$ be the moving-average coefficients
of the VAR in regime $j$. Following Pesaran and Shin (1998), the share of the $H$-step forecast-error variance of variable $i$ due to shock $k$ is
\[
\theta_{ik,j}(H) =
\frac{\sigma_{kk,j}^{-1} \sum_{s=0}^{H-1} (e_i' \Psi_{s,j} \Sigma_j e_k)^2}
{\sum_{s=0}^{H-1} e_i' \Psi_{s,j} \Sigma_j \Psi_{s,j}' e_i},
\]
where $e_i$ is the $n \times 1$ vector with one in position $i$ and zeros elsewhere, $\sigma_{kk,j}$ is the $k$th diagonal element of $\Sigma_j$, and $H$ is the forecast horizon.  Let $\Theta$ be the matrix with $(i,k)$ entry $\theta_{ik,j}(H)$. After each entry of $\Theta$ is divided by its row sum, the total spillover index is defined as the average of the off-diagonal entries in percent. Moreover, for variable $i$, \code{from} is the share
it receives from the other shocks, \code{to} is the share it transmits to
the other variables, and \code{net} is \code{to} minus \code{from}. With \code{boot > 0}, \code{dyspill()} provides bootstrap standard errors
and percentile intervals for the total index and the net spillovers, and for the differences in the
total index between regimes. Within each regime the residuals are
resampled with replacement, the series is rebuilt from the fitted VAR of that regime, and
the model is re-estimated at the same break dates with the same specification.
The function \code{dynobreak()} computes the same measures from a VAR
fitted to the whole sample without breaks, for comparison.

\section{The sysbreak package}
\label{sec:package}

\pkg{sysbreak} is available from the Comprehensive \proglang{R}
Archive Network (CRAN) under the GPL-3 license and can be installed using
\code{install.packages("sysbreak")}. Its page is
\url{https://CRAN.R-project.org/package=sysbreak}. Its only dependency beyond base
\proglang{R} is \pkg{MASS} (Venables and Ripley, 2002), used to compute a
generalized inverse. All results in this paper were produced with
\proglang{R}~4.4.3, \pkg{sysbreak}~1.1.0 and \pkg{MASS}~7.3-64 on
Windows. Below we explain the function \code{sysbreak()}, and the other main functions  \code{varbreak()}, \code{netbreak()}, and \code{ordbreak()} can be applied in the same way. Their arguments, options and default values are explained in the reference manual of the package, available on CRAN.

The function \code{sysbreak()} has the following structure:
\begin{Code}
sysbreak(model, data = NULL, m, coef_breaks = "all", test_level = 0.05,
         ci_level = 0.95, trim = 0.15, verbose = FALSE, brv = FALSE,
         robust = c("none", "hc", "hac"), hetq = FALSE, prewhit = FALSE,
         restart = c("none", "brbp", "brbp_residual"), restart_B = 10,
         restart_verbose = FALSE)
\end{Code}

\begin{sidewaystable}
\centering
\begin{tabular}{@{}p{3.3cm}>{\raggedright\arraybackslash}p{5.6cm}>{\raggedright\arraybackslash}p{8.4cm}>{\raggedright\arraybackslash}p{1.6cm}@{}}
\toprule
\textbf{Argument} & \textbf{Purpose} & \textbf{Options} & \textbf{Default value} \\
\midrule
\code{model} & Specify equations of the system & A formula (same regressors in all equations) or a list of formulas (regressors can differ) & None \\
\code{data} & Data & A data frame & \code{NULL} \\
\code{m} & Maximum number of breaks & A positive integer & None \\
\code{coef\_breaks} & Coefficients allowed to change & \code{"all"}, \code{"none"} (requires \code{brv = TRUE}), or a list of formulas that selects the coefficients & \code{"all"} \\
\code{brv} & Error covariance matrix & \code{FALSE}: constant across regimes. \code{TRUE}: can change & \code{FALSE} \\
\code{test\_level} & Significance level for selecting the number of breaks & \code{0.10}, \code{0.05}, \code{0.025}, or \code{0.01} & \code{0.05} \\
\code{ci\_level} & Coverage rate of the confidence intervals for the break dates & \code{0.95} or \code{0.90} & \code{0.95} \\
\code{trim} & Minimum length of a regime & A fraction of the sample size & \code{0.15} \\
\code{robust} & Heteroskedasticity and serial correlation taken into account for testing, standard errors, and confidence intervals & \code{"none"}: likelihood ratio tests. \code{"hc"}: SupF tests robust to heteroskedasticity. \code{"hac"}: SupF tests robust to heteroskedasticity and serial correlation & \code{"none"} \\
\code{hetq} & Regressor moments for confidence intervals for the break dates & \code{FALSE}: same in all regimes. \code{TRUE}: can differ & \code{FALSE} \\
\code{prewhit} & Prewhitening when \code{robust = "hac"} & \code{FALSE}: none. \code{TRUE}: VAR(1) prewhitening & \code{FALSE} \\
\code{restart} & Bootstrap restart when regression coefficients are subject to restrictions & \code{"none"}, \code{"brbp"} (resampling), or \code{"brbp\_residual"} (residual-based) & \code{"none"} \\
\code{restart\_B} & Number of independent restarts & A positive integer & \code{10} \\
\code{verbose}, \code{restart\_verbose} & Print progress in \proglang{R} console & \code{FALSE} or \code{TRUE} & \code{FALSE} \\
\bottomrule
\end{tabular}
\caption{Arguments of \code{sysbreak()}: purpose, options, and default values.}
\label{tab:arguments}
\end{sidewaystable}

\begin{table}[t!]
\centering
\begin{tabular}{@{}p{4.2cm}>{\raggedright\arraybackslash}p{6.7cm}>{\raggedright\arraybackslash}p{3.4cm}@{}}
\toprule
\textbf{Function} & \textbf{Description} & \textbf{Returns} \\
\midrule
\multicolumn{3}{@{}l}{\textbf{Estimation and testing}} \\[2pt]
\code{sysbreak()} & Breaks in a multivariate regression or a seemingly unrelated regression system & \code{"sysbreak"} object \\
\code{varbreak()} & Breaks in a vector autoregression & \code{"varbreak"} object \\
\code{netbreak()} & Breaks in a network vector autoregression & \code{"netbreak"} object \\
\code{ordbreak()} & Ordered breaks in two subsets of equations & \code{"ordbreak"} object \\
\addlinespace[6pt]
\multicolumn{3}{@{}l}{\textbf{Analysis of a fitted model}} \\[2pt]
\code{brtype()} & Tests of what changes at each break & Data frame \\
\code{dyspill()} & Diebold--Yilmaz spillover measures in each regime, with bootstrap inference & \code{"dyspill"} object \\
\code{dynobreak()} & Diebold--Yilmaz spillover measures from a VAR without breaks & List \\
\addlinespace[6pt]
\multicolumn{3}{@{}l}{\textbf{Methods for summarizing the output}} \\[2pt]
\code{print()}, \code{summary()} & Tests, break dates and coefficient tables & Printed output \\
\code{coef()}, \code{vcov()} & Regime coefficients and their covariance matrix & Vector and matrix, with coefficient names \\
\code{confint()} & Confidence intervals for the break dates or the coefficients & Matrix \\
\code{residuals()}, \code{fitted()} & Residuals and fitted values & Matrices \\
\code{nobs()} & Number of observations used & Integer \\
\code{plot()} & Series with the break dates, their confidence intervals and the fitted values & Figure \\
\bottomrule
\end{tabular}
\caption{Functions of the \pkg{sysbreak} package and methods for the
fitted objects. }
\label{tab:functions}
\end{table}

Table~\ref{tab:arguments} describes the arguments for the function. It can be seen that the inputs to the function are quite intuitive, that is, the user provides formulas for the model
and for the parameters that are allowed to change. Also, the options can
accommodate different numbers of regressors across equations, as well as
a range of situations concerning heteroskedasticity and autocorrelation.

We now turn to the output. A fit from \code{sysbreak()}
is a list with the following: \code{spec} saves the model specification.
\code{testing} saves the test results: the SupLR or SupF
statistics for $k = 1, \ldots, m$ breaks, the
WDmax statistic, and the sequential tests $\text{SEQ}_T(l+1|l)$, with their critical values at the significance level \code{test\_level}.
\code{br\_selected} gives the selected number of breaks. \code{estimation} saves the estimates for the selected
number of breaks: the break dates with their confidence intervals, the
regime coefficients with their covariance matrix, and the error covariance
matrices. The method \code{summary()} prints the tests, the break
dates and the coefficients.
\code{confint()} returns the break dates with their confidence intervals as rows of
the data, and \code{plot()} can be used to draw each series with the break dates, their
confidence intervals and the fitted values.

Table~\ref{tab:functions} summarizes the functions of the
package. It shows that the package is organized around
four estimation functions with the same structure (\code{sysbreak()},
\code{varbreak()}, \code{netbreak()}, and \code{ordbreak()}), two
functions for analyzing a fitted model (\code{brtype()} and
\code{dyspill()}), and the standard \proglang{R} methods for summarizing
the results (\code{summary()}, \code{coef()}, \code{confint()}, and
\code{plot()}), which allows a fit to be handled in the same way as an
\code{lm()} object.

\FloatBarrier
\section{Applications}
\label{sec:applications}

This section presents three applications, one for each of the main
functions \code{sysbreak()}, \code{varbreak()}, and
\code{netbreak()}. All datasets are included in the package. 
\begin{itemize}
\item Application 1 uses \code{sysbreak()} to revisit the effect of house
  prices on the U.S. macroeconomy. It shows how to test for and estimate breaks
  when all coefficients are allowed to change and when only the house
  price coefficients are allowed to change. It also shows how to use the bootstrap restart, and how to
  test for Granger causality within the estimated regimes.
\item Application 2 revisits international inflation dynamics with
  \code{varbreak()}. It first allows breaks in the coefficients only and then allows breaks in both the coefficients and the covariance. It uses
  \code{brtype()} to determine what features of the model change at each break. It uses
  \code{dyspill()} to measure spillovers between countries in each regime,
  with bootstrap inference.
\item Application 3 applies \code{netbreak()} to a network vector
  autoregression with a known network matrix. It shows how to test for and estimate breaks when all coefficients are allowed to change and when only the lag coefficients are allowed to change. 
\end{itemize}

\subsection{Application 1: House prices and the macroeconomy}
\label{sec:app-housing}

Ahamada and Diaz Sanchez (2013) apply the Qu and Perron (2007) procedure
to a vector autoregression in real activity, real house prices, and the
federal funds rate, to study whether the relationship between house prices
and the U.S. macroeconomy has changed over the past fifty years. They find
a break in the mid-1980s and conclude that the effect of house prices on
consumption and output has grown stronger. We revisit their
analysis using \code{sysbreak()}, focusing on output. We first allow all coefficients to change and determine the
number of breaks and their dates. We then allow only the coefficients on
lagged house price growth to change to examine whether the break is in the
house price channel. Finally, we split the sample at the estimated break
date and test for Granger causality between house prices and output within
each regime.

\subsubsection*{Data and model}

After loading the package with \code{library(sysbreak)}, the data can be
loaded using:
\begin{CodeChunk}
\begin{CodeInput}
R> data(housing)
\end{CodeInput}
\end{CodeChunk}
The data can be inspected via:
\begin{CodeChunk}
\begin{CodeInput}
R> head(housing)
\end{CodeInput}
\begin{CodeOutput}
     date         dGDP            dC           dHP  FFR    dI
1 1959:Q4  0.002845754  0.0005456088  0.0030492359 3.98  0.51
2 1960:Q1  0.022237184  0.0099427394  0.0045039917 3.99  0.01
\end{CodeOutput}
\end{CodeChunk}
The dataset contains 200 quarterly observations from 1959:Q4 to 2009:Q3.
The column \code{date} indicates the quarter. The columns \code{dGDP},
\code{dC}, and \code{dHP} are the log differences of real GDP, real
personal consumption expenditures, and real house prices. The column
\code{FFR} is the federal funds rate and \code{dI} is its first
difference. GDP, consumption, the consumer price index, and the funds rate
are from FRED, the database of the Federal Reserve Bank of St. Louis. House prices are the nominal home price index of Shiller
(2015). 

Let $x_t$ denote the log of real GDP, $hp_t$ the log
of real house prices, and $I_t$ the federal funds rate. As in Ahamada
and Diaz Sanchez (2013), the system has three equations with
\[
y_t = (\Delta x_t, \Delta hp_t, \Delta I_t)', \qquad
z_t = (1, \Delta x_{t-1}, \Delta hp_{t-1}, I_{t-1})',
\]
and in regime $j$
\[
y_t = (I_3 \otimes z_t') \beta_j + u_t, \qquad \mathrm{var}(u_t) = \Sigma.
\]
Each regime has 12 coefficients. Note that the funds rate enters the
right-hand side in levels while it is differenced on the left-hand side. The covariance matrix $\Sigma$ is held constant across regimes
(\code{brv = FALSE}). The regressors are constructed as follows:
\begin{CodeChunk}
\begin{CodeInput}
R> n <- nrow(housing)
R> dat <- data.frame(housing[-1, ], L1.dGDP = housing$dGDP[-n],
+                    L1.dHP = housing$dHP[-n], L1.FFR = housing$FFR[-n])
R> dates <- dat$date
R> gdp_model <- cbind(dGDP, dHP, dI) ~ L1.dGDP + L1.dHP + L1.FFR
\end{CodeInput}
\end{CodeChunk}
The last line specifies the model used for estimation. The sample runs from 1960:Q1 to 2009:Q3, with $T = 199$.

\subsubsection*{Step 1: All coefficients allowed to change}

We first allow all 12 coefficients to change. Ahamada and Diaz Sanchez
(2013) use a trimming of 0.20, which is the value used here as well. The
minimum regime length is then $h = 40$ quarters. The model is estimated using the command:
\begin{CodeChunk}
\begin{CodeInput}
R> gdp_unr <- sysbreak(gdp_model, data = dat, m = 3, trim = 0.20)
\end{CodeInput}
\end{CodeChunk}
The omitted arguments take their default values, listed in
Section~\ref{sec:package}: in particular, all coefficients are allowed to change
(\code{coef\_breaks = "all"}), the covariance matrix is held constant
(\code{brv = FALSE}), and the tests are the likelihood-ratio tests
of Section~\ref{sec:methods} rather than their robust versions
(\code{robust = "none"}). The results are summarized using:
\begin{CodeChunk}
\begin{CodeInput}
R> summary(gdp_unr)
\end{CodeInput}
\begin{CodeOutput}
sysbreak: 3 equations, T = 199, m = 3, trim = 0.2, robust = none
Covariance matrix: constant across regimes

Tests for the number of breaks (5% level)
           Statistic Critical value
  WDmax       56.347         31.255 *
  SEQ(2|1)    24.048         32.261
Breaks selected: 1

Break dates (rows of the data) with 95% confidence intervals
          Estimate Lower Upper
  Break 1       92    84   100

Regime 1 (rows 1 to 92)
                   Estimate Std. Error z value Pr(>|z|)
dGDP:(Intercept)  0.0168061  0.0019601   8.574  < 2e-16 ***
dGDP:L1.dGDP     -0.0008881  0.0792066  -0.011  0.99105
dGDP:L1.dHP       0.1270327  0.0760943   1.669  0.09504 .
dGDP:L1.FFR      -0.0012855  0.0002168  -5.928 3.06e-09 ***
...
\end{CodeOutput}
\end{CodeChunk}
The output contains the tests used to select the number of breaks (WDmax and
$\text{SEQ}_T(2|1)$), the
selected number of breaks, the break dates with their confidence intervals, and the
coefficient estimates in each regime.
Here \code{***}, \code{**}, \code{*} and \code{.} indicate statistical significance at the 0.001, 0.01, 0.05 and 0.1 levels, respectively.
For regime 1, only the rows of the GDP equation are shown. The other two
equations and the table for regime 2 (rows 93 to 199) have the same format and
are omitted here. The SupLR statistics for $k = 1, 2, 3$ breaks are stored in
\code{gdp\_unr\$testing}. The WDmax test rejects the null hypothesis of
no break. $\text{SEQ}_T(2|1)$ does not reject,
and therefore one break is selected. The break date and its confidence interval are
reported as rows of \code{dat}; the corresponding quarters can be obtained
using:
\begin{CodeChunk}
\begin{CodeInput}
R> dates[confint(gdp_unr)]
\end{CodeInput}
\begin{CodeOutput}
[1] "1980:Q4" "1982:Q4" "1984:Q4"
\end{CodeOutput}
\end{CodeChunk}
The break date is estimated at 1982:Q4, with a 95\% confidence interval of
[1980:Q4, 1984:Q4].
Ahamada and Diaz
Sanchez (2013)
attribute the break to the transformation of the mortgage market in the
mid-1980s, in particular the growth of secondary markets and the
relaxation of credit constraints, which strengthened the effect of house
prices on the macroeconomy. 

\begin{figure}[h!]
\centering
\includegraphics[width=0.55\textwidth]{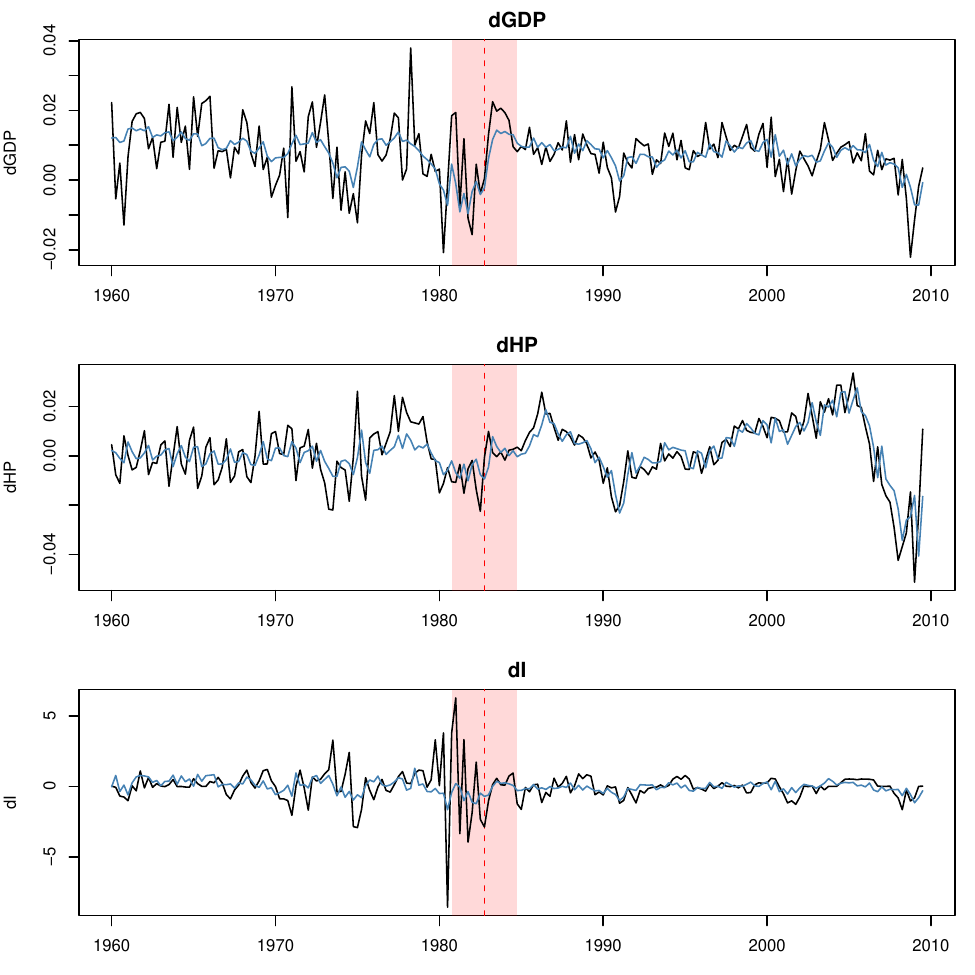}
\caption{Output of \code{plot(gdp\_unr)} for the housing price application with
all coefficients allowed to change: the series (black), the fitted values
(blue), the estimated break date (dashed line) and its 95\% confidence
interval (shaded).}
\label{fig:gdp_unr}
\end{figure}

The fit can be plotted using the following command:
\begin{CodeChunk}
\begin{CodeInput}
R> plot(gdp_unr, dates = seq(1960, by = 0.25, length.out = 199))
\end{CodeInput}
\end{CodeChunk}
Figure~\ref{fig:gdp_unr} shows each series with the estimated break
date, its confidence interval and the fitted values of the dependent variables.

\subsubsection*{Step 2: Only the house price coefficients allowed to change}

The main interest is in the house price channel. It is therefore informative to
allow only the coefficient on $\Delta hp_{t-1}$ to change, in each of the
three equations, and hold the other nine coefficients constant. The
restriction is specified through \code{coef\_breaks} as
\begin{CodeChunk}
\begin{CodeInput}
R> gdp_res <- sysbreak(gdp_model, data = dat, m = 3, trim = 0.20,
+                      coef_breaks = list(dGDP ~ L1.dHP, dHP ~ L1.dHP,
+                                         dI ~ L1.dHP))
\end{CodeInput}
\end{CodeChunk}
The results in this case are obtained using:
\begin{CodeChunk}
\begin{CodeInput}
R> summary(gdp_res)
\end{CodeInput}
\begin{CodeOutput}
sysbreak: 3 equations, T = 199, m = 3, trim = 0.2, robust = none
Covariance matrix: constant across regimes

Tests for the number of breaks (5% level)
           Statistic Critical value
  WDmax       21.547         14.937 *
  SEQ(2|1)     8.758         15.384
Breaks selected: 1

Break dates (rows of the data) with 95% confidence intervals
          Estimate Lower Upper
  Break 1      101    81   121

Coefficients allowed to change: dGDP:L1.dHP, dHP:L1.dHP, dI:L1.dHP

Regime 1 (rows 1 to 101)
                   Estimate Std. Error z value Pr(>|z|)
dGDP:L1.dHP       0.2005574  0.0803040   2.497 0.012508 *
dHP:L1.dHP        0.3482651  0.0904444   3.851 0.000118 ***
dI:L1.dHP        -0.4942391 12.0357816  -0.041 0.967245

Regime 2 (rows 102 to 199)
                   Estimate Std. Error z value Pr(>|z|)
dGDP:L1.dHP       0.1056588  0.0528767   1.998 0.045694 *
dHP:L1.dHP        0.8247614  0.0595538  13.849  < 2e-16 ***
dI:L1.dHP         2.3605649  7.9250493   0.298 0.765809
\end{CodeOutput}
\end{CodeChunk}
Since the WDmax test rejects and
$\text{SEQ}_T(2|1)$ does not, one break is selected. The estimates for the coefficients allowed to change are shown here for
each regime. The other nine coefficients are held constant across regimes
and their rows are omitted. The largest change is in the persistence of house price growth: the
coefficient on $\Delta hp_{t-1}$ in the house price equation rises from
0.35 to 0.82, with standard errors of 0.09 and 0.06. The estimated break date is row 101 of \code{dat}, which is 1985:Q1:
\begin{CodeChunk}
\begin{CodeInput}
R> dates[confint(gdp_res)]
\end{CodeInput}
\begin{CodeOutput}
[1] "1980:Q1" "1985:Q1" "1990:Q1"
\end{CodeOutput}
\end{CodeChunk}
The estimated break date is about 2 years later than in the unrestricted case, though its confidence interval contains the unrestricted estimate. The convergence of the restricted estimator can be examined using the bootstrap restart:
\begin{CodeChunk}
\begin{CodeInput}
R> set.seed(1)
R> gdp_res_b <- sysbreak(gdp_model, data = dat, m = 3, trim = 0.20,
+                        coef_breaks = list(dGDP ~ L1.dHP, dHP ~ L1.dHP,
+                                           dI ~ L1.dHP), restart = "brbp")
R> dates[gdp_res_b$estimation$breaks]
\end{CodeInput}
\begin{CodeOutput}
[1] "1985:Q1"
\end{CodeOutput}
\end{CodeChunk}
The estimated break date is unchanged.

\subsubsection*{Step 3: Granger causality within each regime}

We split the sample at the break date of the restricted model and fit an
unrestricted VAR(1) within each regime. This is done using
\code{granger()}, a function defined in the replication script that accompanies this paper:
\begin{CodeChunk}
\begin{CodeInput}
R> granger(gdp_res, "dGDP")
\end{CodeInput}
\begin{CodeOutput}
           Regime hp_to_x     p x_to_hp     p
1 1960:Q1-1985:Q1    2.91 0.088    0.39 0.534
2 1985:Q2-2009:Q3    8.04 0.005    5.34 0.021
\end{CodeOutput}
\end{CodeChunk}
The column \code{hp\_to\_x} tests whether house prices Granger-cause GDP, and \code{x\_to\_hp} the reverse. The columns \code{p} report the corresponding $p$-values, denoted by $p$ below. Before the break,
house prices do not Granger-cause GDP at the 5\% significance level ($p = 0.088$). After the break they do, with $p = 0.005$. This agrees with the finding of Ahamada and
Diaz Sanchez (2013) that the effect of house prices on activity has become
stronger since the mid-1980s. In the reverse direction, GDP does not Granger-cause house prices in
the first regime but does in the second ($p = 0.021$). This reinforces the conclusion.
The results are similar if the sample is split at the break date of the
unrestricted model, 1982:Q4, using \code{granger(gdp\_unr, "dGDP")}: the
$p$-values for house prices Granger-causing GDP are 0.184 before the break
and 0.011 after, and those for the reverse direction are 0.487 and 0.040.

\subsection{Application 2: International inflation dynamics}
\label{sec:app-inflation}

Bataa et al.~(2013) study whether the linkage between
inflation rates in Canada, the euro area, the United Kingdom, and the United
States has changed since the 1970s. They estimate a VAR and date the
breaks in its coefficients and in its error covariance separately. We revisit this
exercise using \code{varbreak()}, with common break dates for the
coefficients and the error covariance. We then examine what changes at each break
and measure how the spillovers between countries differ across regimes.

\subsubsection*{Data and model}

The data can be loaded and inspected using:
\begin{CodeChunk}
\begin{CodeInput}
R> data(inflation)
R> print(head(inflation), digits = 3)
\end{CodeInput}
\begin{CodeOutput}
     date Canada EuroArea France Germany Italy Japan    UK    US
1 1973:04  1.461    0.569  0.475   0.463 1.045 0.952 0.477 0.673
2 1973:05  0.179    0.892  0.968   0.743 1.580 1.478 0.905 0.599
\end{CodeOutput}
\end{CodeChunk}
The dataset contains 417 monthly observations from 1973:04 to 2007:12 for
the G7 countries and the euro area. The series are those of Bataa
et al.~(2013), who adjusted them for outliers and seasonality. Let $y_t$ collect the inflation rates of Canada, the United Kingdom, the
euro area, and the United States. In regime $j$,
\[
y_t = c_j + A_j y_{t-1} + u_t, \qquad \mathrm{var}(u_t) = \Sigma_j .
\]
Each regime has 20 coefficients. The data are converted to a matrix using:
\begin{CodeChunk}
\begin{CodeInput}
R> intl  <- as.matrix(inflation[, c("Canada", "UK", "EuroArea", "US")])
R> dates <- inflation$date
\end{CodeInput}
\end{CodeChunk}

\subsubsection*{Step 1: Breaks in coefficients with a constant covariance}

We first allow all 20 coefficients to change while holding $\Sigma$ constant.
As in Bataa et al.~(2013), the trimming is 0.15 and at most five breaks are
permitted. The minimum regime length is then $h = 62$ months. The model is
estimated using:
\begin{CodeChunk}
\begin{CodeInput}
R> intl_coef <- varbreak(intl, lags = 1, m = 5, trim = 0.15)
\end{CodeInput}
\end{CodeChunk}
The omitted arguments take their default values: an intercept is
included, all coefficients may change, and $\Sigma$ is constant
(\code{brv = FALSE}).
The results are summarized using:
\begin{CodeChunk}
\begin{CodeInput}
R> summary(intl_coef)
\end{CodeInput}
\begin{CodeOutput}
varbreak: 4 equations, T = 416, m = 5, trim = 0.15, robust = none
Covariance matrix: constant across regimes

Tests for the number of breaks (5% level)
           Statistic Critical value
  WDmax      143.099         44.636 *
  SEQ(2|1)    80.765         45.618 *
  SEQ(3|2)    35.737         46.998
Breaks selected: 2

Break dates (rows of the data) with 95% confidence intervals
          Estimate Lower Upper
  Break 1      125   121   129
  Break 2      220   215   225
\end{CodeOutput}
\end{CodeChunk}
The tables for the estimated coefficients are omitted
here. The WDmax test rejects the null hypothesis of no break.
$\text{SEQ}_T(2|1)$ rejects while $\text{SEQ}_T(3|2)$ does not, and therefore two
breaks are selected. The break dates and their confidence intervals, as
months, are:
\begin{CodeChunk}
\begin{CodeInput}
R> ci <- confint(intl_coef); ci[] <- dates[ci]; noquote(ci)
\end{CodeInput}
\begin{CodeOutput}
       2.5 %   break   97.5 %
break1 1983:04 1983:08 1983:12
break2 1991:02 1991:07 1991:12
\end{CodeOutput}
\end{CodeChunk}
Bataa et al.~(2013) find the same number of coefficient breaks, in 1982:06
and 1990:10. They estimate the coefficient breaks by feasible GLS, conditioning
on separate covariance breaks. Such conditioning is not used here.

This step will return a warning that no segment has enough observations
to insert an additional break. This happens because after four breaks have
been inserted, the resulting segments are all shorter than 124 months, twice
the minimum length for a regime. This warning signals that many breaks are
allowed relative to the sample size and is not an error. The same warning
appears in Step~2.

\subsubsection*{Step 2: Common breaks in the coefficients and the covariance}

\begin{figure}[t!]
\centering
\includegraphics[width=0.7\textwidth]{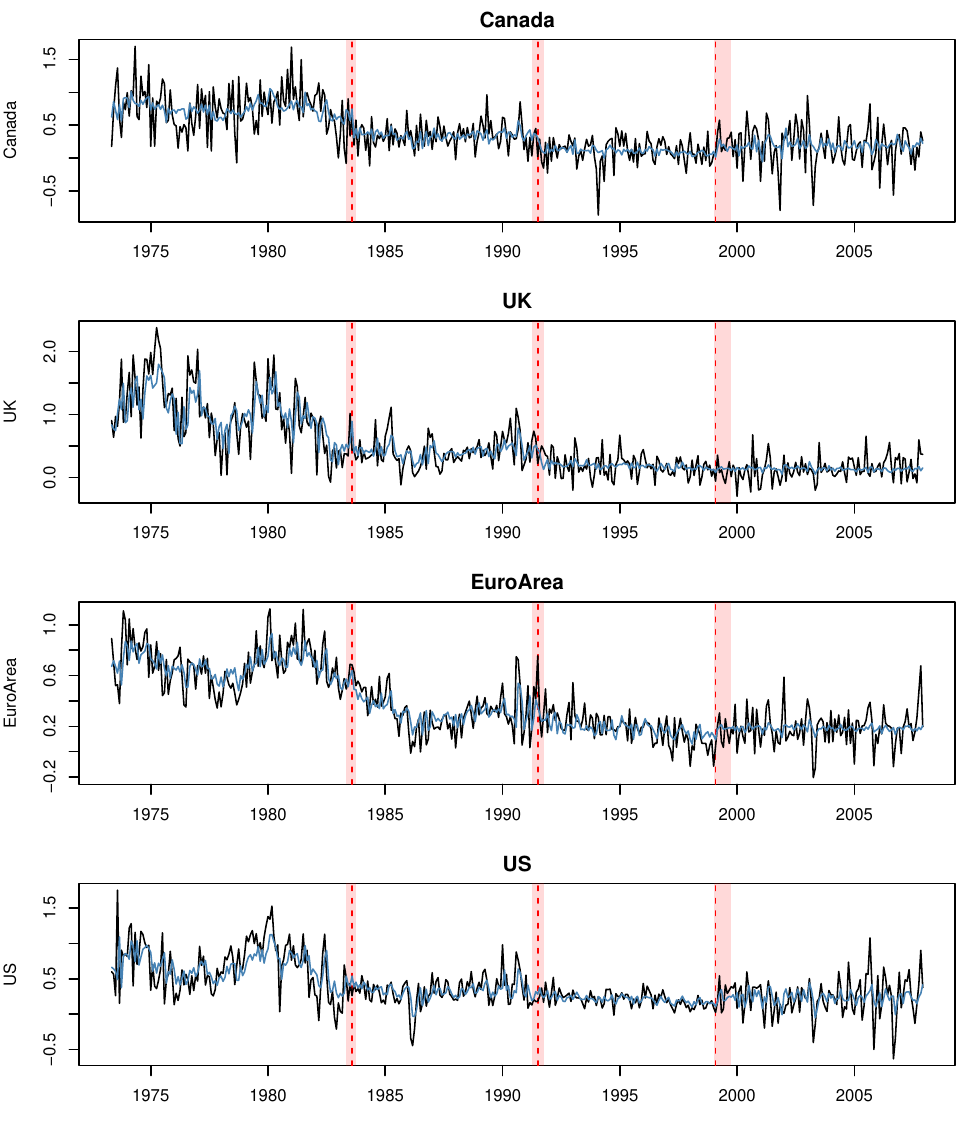}
\caption{Output of \code{plot(intl\_cov)} for the inflation application: the four inflation
series (black), the fitted values (blue), the three estimated break dates
(dashed lines) and their 95\% confidence intervals (shaded).}
\label{fig:intl_cov}
\end{figure}

We next allow $\Sigma$ to change at the same dates as the coefficients:
\begin{CodeChunk}
\begin{CodeInput}
R> intl_cov <- varbreak(intl, lags = 1, m = 5, trim = 0.15, brv = TRUE)
R> summary(intl_cov)
\end{CodeInput}
\begin{CodeOutput}
varbreak: 4 equations, T = 416, m = 5, trim = 0.15, robust = none
Covariance matrix: may change across regimes

Tests for the number of breaks (5% level)
           Statistic Critical value
  WDmax      327.906         59.118 *
  SEQ(2|1)   111.807         60.252 *
  SEQ(3|2)   125.968         61.730 *
  SEQ(4|3)    56.498         62.632
Breaks selected: 3

Break dates (rows of the data) with 95% confidence intervals
          Estimate Lower Upper
  Break 1      125   122   127
  Break 2      220   217   223
  Break 3      311   310   319
\end{CodeOutput}
\begin{CodeInput}
R> ci <- confint(intl_cov); ci[] <- dates[ci]; noquote(ci)
\end{CodeInput}
\begin{CodeOutput}
       2.5 %   break   97.5 %
break1 1983:05 1983:08 1983:10
break2 1991:04 1991:07 1991:10
break3 1999:01 1999:02 1999:10
\end{CodeOutput}
\end{CodeChunk}
Now three breaks are selected. The first two are the same as those found in Step~1; the third, 1999:02, is new. Step~3 below will show that at this break, the error covariance changes while the
coefficients remain constant. Figure~\ref{fig:intl_cov} plots the fit:
\begin{CodeChunk}
\begin{CodeInput}
R> plot(intl_cov, dates = seq(1973 + 3/12, by = 1/12, length.out = 417))
\end{CodeInput}
\end{CodeChunk}
The 1983 break coincides with a fall in both the level and the
volatility of inflation in all four countries. The 1999 break is not
visible in the individual series. Below this is studied further.

\subsubsection*{Step 3: What changes at each break}

The function \code{brtype()} examines whether the
volatilities, the correlations, and the coefficients differ between the two
regimes. The volatility statistic is a two-sample
Hotelling statistic on the squared residuals, with $n = 4$ degrees of
freedom. The correlation statistic is the test of Jennrich (1970), with
$n(n-1)/2 = 6$ degrees of freedom. The coefficient statistic is a Wald
test with 20 degrees of freedom. The Wald test holds the error covariance constant across
regimes, because estimating it separately in each regime makes the test
oversized in samples of this length.
\begin{CodeChunk}
\begin{CodeInput}
R> types <- brtype(intl_cov)
R> out <- t(round(types[-1], 4)); colnames(out) <- dates[types$breaks]; out
\end{CodeInput}
\begin{CodeOutput}
               1983:08 1991:07 1999:02
volatility     49.7829 13.3446 20.4555
volatility_p    0.0000  0.0097  0.0004
correlation     6.5257  5.8244 23.3984
correlation_p   0.3669  0.4431  0.0007
coefficients   67.3761 57.3295 22.5723
coefficients_p  0.0000  0.0000  0.3103
\end{CodeOutput}
\end{CodeChunk}

For each break, the test statistic (e.g., \code{volatility}) and its $p$-value (\code{volatility\_p}) are reported. The results suggest that the first two breaks change the volatilities and the coefficients but not
the correlations. The third break changes the volatilities and the correlations
but not the coefficients. Bataa et al.~(2013) also find volatility changes
at all three of their covariance breaks, with a significant correlation break only in 1999.

Furthermore, for each regime, the errors' standard deviations and their correlation with other countries can be computed as follows. Note that
\code{intl\_cov\$estimation\$covariance} is a $16 \times 4$ matrix, with the
four covariance matrices arranged on top of each other.
\begin{CodeChunk}
\begin{CodeInput}
R> cov_r <- lapply(1:4, function(r)
+    intl_cov$estimation$covariance[(r - 1) * 4 + 1:4, ])
R> sds <- sapply(cov_r, function(s) sqrt(diag(s)))
R> dimnames(sds) <- list(colnames(intl), paste("regime", 1:4))
R> round(sds, 3)
\end{CodeInput}
\begin{CodeOutput}
         regime 1 regime 2 regime 3 regime 4
Canada      0.326    0.178    0.196    0.293
UK          0.396    0.216    0.175    0.184
EuroArea    0.150    0.143    0.104    0.136
US          0.308    0.178    0.088    0.249
\end{CodeOutput}
\begin{CodeInput}
R> round(sds[, -1] / sds[, -4], 2)
\end{CodeInput}
\begin{CodeOutput}
         regime 2 regime 3 regime 4
Canada       0.54     1.10     1.50
UK           0.54     0.81     1.05
EuroArea     0.95     0.73     1.31
US           0.58     0.50     2.82
\end{CodeOutput}
\begin{CodeInput}
R> round(sapply(cov_r, function(s) mean(cov2cor(s)[upper.tri(s)])), 3)
\end{CodeInput}
\begin{CodeOutput}
[1] 0.075 0.144 0.140 0.450
\end{CodeOutput}
\end{CodeChunk}
In 1983 the standard deviations for Canada, the United Kingdom, and the
United States fall by 46\%, 46\%, and 42\%, respectively, while that of the euro area barely
changes. In 1999 the volatility rises in Canada, the euro area, and the United
States, but not in the United Kingdom. The mean correlation is between
0.075 and 0.144 in the first three regimes and 0.450 in the last. This suggests that the correlation is much stronger in the last regime, which is further studied below through spillover measures.

\subsubsection*{Step 4: Spillovers}

The function \code{dyspill()} computes the spillover measures of
Diebold and Yilmaz (2012) in each regime. For a 12-month horizon, the measures
are computed using:
\begin{CodeChunk}
\begin{CodeInput}
R> spill <- dyspill(intl_cov, h = 12)
R> spill
\end{CodeInput}
\begin{CodeOutput}
Diebold-Yilmaz generalized spillover measures
Model: varbreak | Horizon: 12
 regime  sample spectral_radius stable total benchmark
      1   2:125          0.7286   TRUE 17.83      8.12
      2 126:220          0.4483   TRUE 22.09     15.05
      3 221:311          0.4582   TRUE 19.34      9.46
      4 312:417          0.2675   TRUE 40.87     39.06
\end{CodeOutput}
\end{CodeChunk}
The column \code{sample} gives the first and last observations of each
regime; for example, regime 1 runs from row 2 to row 125. \code{spectral\_radius} measures the stationarity of the VAR in each
regime; the VAR is stationary when its value is below one, which is the case in all
four regimes. The total index, reported in the column
\code{total}, is the share of the
forecast-error variance due to shocks from other countries, averaged over
countries and expressed in percent. The index is about 20 in the first three regimes and doubles in
1999. Before 1999, about a fifth of the forecast-error variance of a
country's inflation over a year is attributed to the innovations of the
other three countries. After 1999, the share is about two fifths. Note that the generalized
decomposition does not orthogonalize the innovations, so this is an
attribution, not an identification of where the shocks originate.

The column \code{benchmark}
is the total index computed with $A_j$ set to zero. It measures the
spillover due to the correlation of the innovations only, with no lagged
transmission. The ratio of the benchmark to the total index shows how much of the
spillover is simultaneous rather than transmitted with a lag through
$A_j$:
\begin{CodeChunk}
\begin{CodeInput}
R> total     <- sapply(spill$regimes, function(x) x$total)
R> benchmark <- sapply(spill$regimes, function(x) x$benchmark_total)
R> round(100 * benchmark / total, 1)
\end{CodeInput}
\begin{CodeOutput}
[1] 45.6 68.1 48.9 95.6
\end{CodeOutput}
\end{CodeChunk}
In the last regime the error correlations account for almost all of the
index. Therefore, after 1999
the spillover is almost entirely simultaneous. In the first and third regimes they account for about half of the
index, and the remaining part is due to the dynamics.  This agrees with Step~3, where
the 1999 break changes the correlations but not the coefficients.

The directional measures \code{from}, \code{to} and \code{net}
of each country are stored in \code{spill\$regimes[[r]]} for regime
\code{r}; for example, \code{sapply(spill\$regimes, function(x)
x\$net)} gives the net positions of the four countries in the four
regimes. They are omitted here. 

The function \code{dynobreak()} computes the same measures from a VAR
fitted to the whole sample, without breaks:
\begin{CodeChunk}
\begin{CodeInput}
R> round(dynobreak(intl, lags = 1, h = 12)$total, 2)
\end{CodeInput}
\begin{CodeOutput}
[1] 47.38
\end{CodeOutput}
\end{CodeChunk}
The index without breaks exceeds that of every regime, including the last.
Therefore, ignoring the breaks would overstate the spillovers throughout the sample.

\subsubsection*{Step 5: Is the spillover index constant before 1999?}

With \code{boot > 0}, \code{dyspill()} produces bootstrap standard errors
and percentile intervals. Within each regime, we sample the residuals with replacement.
The series is then rebuilt using the fitted VAR of that regime, and the model is
re-estimated. The break dates are treated as known:
\begin{CodeChunk}
\begin{CodeInput}
R> spill_b <- dyspill(intl_cov, h = 12, boot = 1000, seed = 20260814)
R> spill_b
\end{CodeInput}
\begin{CodeOutput}
Diebold-Yilmaz generalized spillover measures
Model: varbreak | Horizon: 12
 regime  sample spectral_radius stable total benchmark   se lower upper
      1   2:125          0.7286   TRUE 17.83      8.12 4.71 12.08 30.27
      2 126:220          0.4483   TRUE 22.09     15.05 6.71 12.24 37.47
      3 221:311          0.4582   TRUE 19.34      9.46 4.26 15.27 31.37
      4 312:417          0.2675   TRUE 40.87     39.06 3.83 33.75 48.62

Bootstrap: 1000 replications, 95% percentile intervals
Differences in the total index between regimes:
 regimes difference   se  lower upper
     2-1       4.27 8.36 -11.07 20.51
     3-1       1.52 6.37 -10.98 14.91
     4-1      23.04 6.06   8.44 32.17
     3-2      -2.75 7.89 -17.37 13.03
     4-2      18.78 7.83   1.92 31.95
     4-3      21.52 5.86   7.15 29.65
\end{CodeOutput}
\end{CodeChunk}
Note that the confidence intervals for the differences among the first three regimes all
contain zero, while those between the last regime and each of the others exclude zero.
A Wald test that the index is the same in the first three regimes can be implemented using:
\begin{CodeChunk}
\begin{CodeInput}
R> d    <- spill_b$boot$total[, 2:3] - spill_b$boot$total[, 1]
R> dobs <- total[2:3] - total[1]
R> wald <- drop(t(dobs) %*% solve(cov(d), dobs))
R> c(Wald = round(wald, 3),
+    p = round(pchisq(wald, 2, lower.tail = FALSE), 3))
\end{CodeInput}
\begin{CodeOutput}
 Wald     p
0.261 0.878
\end{CodeOutput}
\end{CodeChunk}
The test does not reject the null hypothesis that the index is the same in
the first three regimes. The data are consistent with an index of about 20
until 1999 and about 41 afterwards. Thus, the change in the index is
associated with the change in the correlations in 1999, and not with the
changes in the volatilities and the dynamics in 1983 and 1991. This analysis
shows that the procedures not only detect and date the breaks but also help trace the nature of the changes through
\code{brtype()} and \code{dyspill()}.

\subsection{Application 3: A network vector autoregression}
\label{sec:app-network}

Zhu et al.~(2017) proposed the network vector
autoregression, where each unit (i.e., each time series) responds to the average of its linked units and to its own past. A specification with $p$ lags and unit-specific intercepts is:
\[
y_t = \mu + \sum_{\ell=1}^{p} (\alpha_{\ell} W + \gamma_{\ell} I)\, y_{t-\ell} + u_t,
\]
where $y_t = (y_{1,t}, \ldots, y_{n,t})'$ contains the $n$ units, $\mu$ is the
vector of intercepts, and $W$ is the $n \times n$ network matrix. Its $(i, l)$ entry 
is $1/n_i$ if unit $i$ is linked to unit $l$ and 0 otherwise, where $n_i$ is the
number of units that $i$ is linked to. The scalars
$\alpha_{\ell}$ and $\gamma_{\ell}$ capture the dependence of each unit on this
network average and on its own lags, respectively. Han and Lee (2024) study structural breaks in this model and select the breaks by a
minimum description length criterion. The current package provides an alternative approach
based on quasi-maximum likelihood, implemented in \code{netbreak()}. To illustrate
the method, we use a simulated dataset from the data generating process of
Scenario 1 in Han and Lee (2024), described below.

\subsubsection*{Data and model}

The data have 300 observations on $n = 20$ units, with breaks occurring at $t=100$
and $t=200$. In regime $j$ ($j = 1, 2, 3$), the data are generated by
\[
y_t = \mu_j + \sum_{\ell=1}^{2} (\alpha_{\ell,j} W + \gamma_{\ell,j} I)
y_{t-\ell} + u_t.
\]
In the three regimes, the values of
$(\alpha_1, \alpha_2, \gamma_1, \gamma_2)$ are $(-0.10, 0, 0.20, 0)$,
$(0.20, -0.22, -0.12, 0.40)$ and $(-0.12, 0.10, 0.25, -0.40)$. The intercepts
$\mu_j$ also change across regimes. As in Han and Lee (2024), they are generated as $\mu_j = 0.15 Z \delta_j$, where the values of $\delta_j$ are $(0.1, 0.4, 0.1, 0.2)'$, $(-0.1, 0.1, 0.2, -0.1)'$ and $(0.2, -0.5, 0.1, 0.1)'$. The rows of the $n \times 4$ matrix $Z$ are drawn independently from $N(0, \Sigma_Z)$, where the $(a, b)$ entry of $\Sigma_Z$ is $0.5^{|a-b|}$.
Each regime thus has 24 coefficients. The error variance is $0.01$ in all
regimes. The network $W$ is generated as in Zhu et al.~(2017): for each unit, the
number of units linked to it, denoted by $x$, is drawn from $\{1, \ldots, 19\}$ with
probability proportional to $x^{-1.2}$, so that a few units receive many links
and most receive few. The simulated dataset is included in the package as \code{netsim} and can be loaded using:
\begin{CodeChunk}
\begin{CodeInput}
R> data(netsim)
R> y <- netsim$y
R> W <- netsim$W
R> dim(y)
\end{CodeInput}
\begin{CodeOutput}
[1] 300  20
\end{CodeOutput}
\end{CodeChunk}

We take two steps. We first allow all 24 coefficients to change, with the covariance matrix held constant. We then
allow only the four lag coefficients to change and hold the intercepts
constant. This restriction is false since the intercepts are also affected by the breaks. Therefore, the second step illustrates the effect of false
restrictions on the estimated break dates and coefficients.

\subsubsection*{Step 1: All coefficients allowed to change}

The model can be estimated using:
\begin{CodeChunk}
\begin{CodeInput}
R> fit_all <- netbreak(y, W, lags = 2, m = 5, trim = 0.15)
\end{CodeInput}
\end{CodeChunk}
The results are:
\begin{CodeChunk}
\begin{CodeInput}
R> summary(fit_all)
\end{CodeInput}
\begin{CodeOutput}
netbreak: 20 equations, T = 298, m = 5, trim = 0.15, robust = none
Covariance matrix: constant across regimes

Tests for the number of breaks (5% level)
           Statistic Critical value
  WDmax     1143.716         50.533 *
  SEQ(2|1)   514.749         51.592 *
  SEQ(3|2)    30.736         53.018
Breaks selected: 2

Break dates (rows of the data) with 95% confidence intervals
          Estimate Lower Upper
  Break 1      100    99   101
  Break 2      200   199   201
\end{CodeOutput}
\end{CodeChunk}
The WDmax test
rejects the null hypothesis of no breaks. $\text{SEQ}_T(2|1)$ rejects, while $\text{SEQ}_T(3|2)$ does not, and therefore 
two breaks are selected. These are estimated at rows 100 and 200, i.e., the true break dates, with
confidence intervals of one observation on each side. The estimation results for the coefficients are omitted to save space. 

\begin{figure}[h!]
\centering
\includegraphics[width=0.8\textwidth]{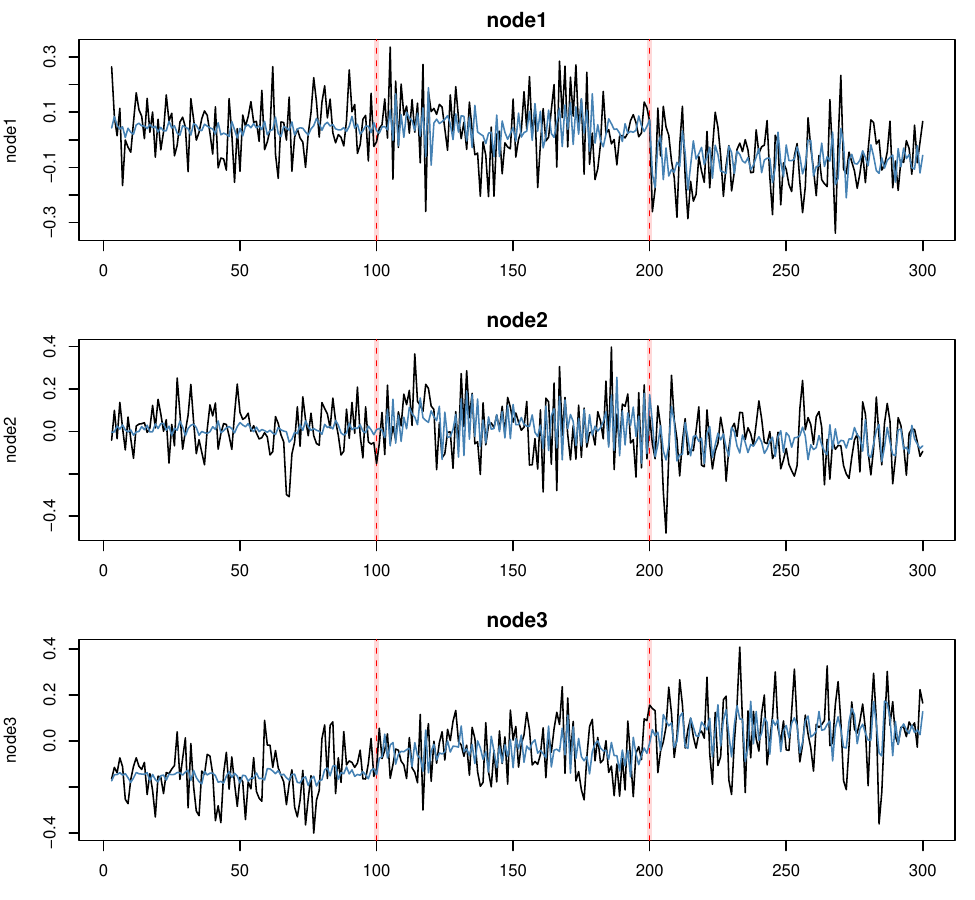}
\caption{Output of \code{plot(fit\_all, which = 1:3)} in the network VAR application: the first
three units (black), the fitted values (blue) and the two estimated break
dates (dashed).}
\label{fig:fit_all}
\end{figure}

The argument \code{which} selects the series to plot.
Figure~\ref{fig:fit_all} shows the first three
units, obtained using:
\begin{CodeChunk}
\begin{CodeInput}
R> plot(fit_all, which = 1:3)
\end{CodeInput}
\end{CodeChunk}
In the figure, the confidence intervals are too narrow to be visible.

This step will return a warning that \code{trim} may be too small. This
happens because the minimum length for a regime is small relative to the
dimension of the system (the number of equations and the number of
regressors in each equation). The warning signals that, in such a situation,
the estimated break dates tend to produce regimes close to the minimum
length. It is not an error. The same warning appears in Step~2.

\subsubsection*{Step 2: Only the lag coefficients allowed to change}

The argument \code{coef\_breaks} can be
used to choose which parameter groups are allowed to change. These can be
\code{"intercepts"}, \code{"network"} for $\alpha_1$ and $\alpha_2$,
\code{"own"} for $\gamma_1$ and $\gamma_2$,
\code{"dynamic"} for all lag coefficients, and \code{"lag1"} or
\code{"lag2"} for $\alpha_\ell$ and $\gamma_\ell$ at that lag. We use \code{"dynamic"}:
\begin{CodeChunk}
\begin{CodeInput}
R> fit_dyn <- netbreak(y, W, lags = 2, m = 5, trim = 0.15,
+                      coef_breaks = "dynamic")
R> summary(fit_dyn)
\end{CodeInput}
\begin{CodeOutput}
netbreak: 20 equations, T = 298, m = 5, trim = 0.15, robust = none
Covariance matrix: constant across regimes

Tests for the number of breaks (5% level)
           Statistic Critical value
  WDmax      767.405         17.831 *
  SEQ(2|1)   279.555         18.144 *
  SEQ(3|2)     7.539         19.033
Breaks selected: 2

Break dates (rows of the data) with 95% confidence intervals
          Estimate Lower Upper
  Break 1      100    98   102
  Break 2      200   199   201

\end{CodeOutput}
\end{CodeChunk}
The same two breaks are selected, and both are estimated at the true dates. The confidence interval for the first break is now
  $[98, 102]$, compared with $[99, 101]$ in Step 1. The coefficient estimates, however, are affected by the restriction. In regime 1,
  the estimates of $\gamma_1$ and $\gamma_2$ are 0.33 and 0.15, each with a standard error of 0.02, while the true values are 0.20
  and 0. Thus, in this example, holding the intercepts constant has little effect on the estimated break dates, but the estimates of
  the lag coefficients are biased away from their true values.

\paragraph{Computational time.} The three applications take approximately 2,
7 and 14 minutes on a desktop computer, with \proglang{R}~4.4.3 and
\pkg{sysbreak}~1.1.0 on Windows. Each application estimates more than
one specification, so the time for a single estimation is shorter.

\section{Conclusion}
\label{sec:conclusion}

This paper introduced the \pkg{sysbreak} package, which implements a
set of methods for estimating and testing multiple structural breaks in
linear systems of equations. The methods cover multivariate regressions,
seemingly unrelated regressions, vector autoregressions, and network
vector autoregressions. They allow breaks in a subset of the
coefficients, in the error covariance matrix, or in both, and provide
likelihood-ratio and HAC-robust tests for selecting
the number of breaks, as well as confidence intervals for the break
dates and the coefficients. Two additional functions can be used to analyze a fitted
model: \code{brtype()} tests whether the volatilities, the
correlations, and the coefficients change at each break, and
\code{dyspill()} computes regime-specific spillover measures with
bootstrap inference, for applications in which the propagation of shocks
across equations is of interest. We hope this package
provides a useful foundation for researchers working with multivariate
models in environments where structural stability cannot be taken for
granted.

\linespread{1}\selectfont
\section*{References}

\begin{list}{}{\setlength{\leftmargin}{2em}\setlength{\itemindent}{-2em}}

\item Ahamada, I., and Diaz Sanchez, J.\ L.\ (2013). A Retrospective
Analysis of the House Price Macro-Relationship in the United States.
\emph{International Journal of Central Banking}, 9, 153--174.

\item Anderson, T.\ W.\ (2003). \emph{An Introduction to Multivariate
Statistical Analysis}, 3rd ed. Wiley, Hoboken.

\item Andrews, D.\ W.\ K.\ (1991). Heteroskedasticity and Autocorrelation
Consistent Covariance Matrix Estimation. \emph{Econometrica}, 59, 817--858.

\item Bai, J.\ (1997). Estimation of a Change Point in Multiple Regression
Models. \emph{Review of Economics and Statistics}, 79, 551--563.

\item Bai, J.\ (2000). Vector Autoregressive Models with Structural Changes
in Regression Coefficients and in Variance--Covariance Matrices.
\emph{Annals of Economics and Finance}, 1, 303--339.

\item Bai, J., Lumsdaine, R.\ L., and Stock, J.\ H.\ (1998). Testing for
and Dating Common Breaks in Multivariate Time Series. \emph{Review of
Economic Studies}, 65, 395--432.

\item Bai, J., and Perron, P.\ (1998). Estimating and Testing Linear Models
with Multiple Structural Changes. \emph{Econometrica}, 66, 47--78.

\item Bai, J., and Perron, P.\ (2003a). Computation and Analysis of Multiple
Structural Change Models. \emph{Journal of Applied Econometrics}, 18, 1--22.

\item Bai, J., and Perron, P.\ (2003b). Critical Values for Multiple
Structural Change Tests. \emph{Econometrics Journal}, 6, 72--78.

\item Bartlett, M.\ S.\ (1938). Further Aspects of the Theory of Multiple
Regression. \emph{Proceedings of the Cambridge Philosophical Society}, 34,
33--40.

\item Bataa, E., Osborn, D.\ R., Sensier, M., and van Dijk, D.\ (2013).
Structural Breaks in the International Dynamics of Inflation. \emph{Review
of Economics and Statistics}, 95, 646--659.

\item Blatt, D., Candelon, B., and Manner, H.\ (2015). Detecting Contagion
in a Multivariate Time Series System: An Application to Sovereign Bond
Markets in Europe. \emph{Journal of Banking \& Finance}, 59, 1--13.

\item Breitung, J., and Eickmeier, S.\ (2011). Testing for Structural
Breaks in Dynamic Factor Models. \emph{Journal of Econometrics}, 163,
71--84.

\item Chen, L., Dolado, J.\ J., and Gonzalo, J.\ (2014). Detecting Big
Structural Breaks in Large Factor Models. \emph{Journal of Econometrics},
180, 30--48.

\item Diebold, F.\ X., and Yilmaz, K.\ (2012). Better to Give than to
Receive: Predictive Directional Measurement of Volatility Spillovers.
\emph{International Journal of Forecasting}, 28, 57--66.

\item Han, X., and Inoue, A.\ (2015). Tests for Parameter Instability in
Dynamic Factor Models. \emph{Econometric Theory}, 31, 1117--1152.

\item Han, Y., and Lee, T.\ C.\ M.\ (2024). Structural Break Detection in
Non-Stationary Network Vector Autoregression Models. \emph{IEEE
Transactions on Network Science and Engineering}, 11, 4134--4145.

\item James, N.\ A., and Matteson, D.\ S.\ (2015). \pkg{ecp}: An
\proglang{R} Package for Nonparametric Multiple Change Point Analysis of
Multivariate Data. \emph{Journal of Statistical Software}, 62(7), 1--25.

\item Jennrich, R.\ I.\ (1970). An Asymptotic $\chi^2$ Test for the Equality
of Two Correlation Matrices. \emph{Journal of the American Statistical
Association}, 65, 904--912.

\item Li, D., Qian, J., and Su, L.\ (2016). Panel Data Models With
Interactive Fixed Effects and Multiple Structural Breaks. \emph{Journal of
the American Statistical Association}, 111, 1804--1819.

\item Nguyen, L., Yamamoto, Y., and Perron, P.\ (2024). \pkg{mbreaks}:
Estimation and Inference for Structural Breaks in Linear Regression Models.
\proglang{R} package version 1.0.1,
\url{https://CRAN.R-project.org/package=mbreaks}.

\item Okui, R., and Wang, W.\ (2021). Heterogeneous Structural Breaks in
Panel Data Models. \emph{Journal of Econometrics}, 220, 447--473.

\item Pesaran, H.\ H., and Shin, Y.\ (1998). Generalized Impulse Response
Analysis in Linear Multivariate Models. \emph{Economics Letters}, 58,
17--29.

\item Qian, J., and Su, L.\ (2016). Shrinkage Estimation of Common Breaks
in Panel Data Models via Adaptive Group Fused Lasso. \emph{Journal of
Econometrics}, 191, 86--109.

\item Qu, Z., and Perron, P.\ (2007). Estimating and Testing Structural
Changes in Multivariate Regressions. \emph{Econometrica}, 75, 459--502.

\item R Core Team (2025). \proglang{R}: A Language and Environment for
Statistical Computing. \proglang{R} Foundation for Statistical Computing,
Vienna, Austria. \url{https://www.R-project.org/}.

\item Shiller, R.\ J.\ (2015). \emph{Irrational Exuberance}, 3rd ed.
Princeton: Princeton University Press.

\item Tobing, H., and McGilchrist, C.\ A.\ (1992). Recursive Residuals for
Multivariate Regression Models. \emph{Australian Journal of Statistics},
34, 217--232.

\item Venables, W.\ N., and Ripley, B.\ D.\ (2002). \emph{Modern Applied
Statistics with S}, 4th ed. Springer, New York.

\item Wood, S.\ N.\ (2001). Minimizing Model Fitting Objectives That Contain
Spurious Local Minima by Bootstrap Restarting. \emph{Biometrics}, 57,
240--244.

\item Zeileis, A., Leisch, F., Hornik, K., and Kleiber, C.\ (2002).
\pkg{strucchange}: An \proglang{R} Package for Testing for Structural
Change in Linear Regression Models. \emph{Journal of Statistical
Software}, 7(2), 1--38.

\item Zeileis, A., Kleiber, C., Kr\"amer, W., and Hornik, K.\ (2003).
Testing and Dating of Structural Changes in Practice. \emph{Computational
Statistics \& Data Analysis}, 44, 109--123.

\item Zhu, X., Pan, R., Li, G., Liu, Y., and Wang, H.\ (2017). Network
Vector Autoregression. \emph{Annals of Statistics}, 45, 1096--1123.

\end{list}

\end{document}